# LED-Induced Fluorescence System for Tea Classification and Quality Assessment

Running title：LED-Induced Fluorescence for Tea Quality Assessment

Yongjiang Dong, Xuan Liu, Liang Mei, Chao Feng, Chunsheng Yan, Sailing He*

Centre for Optical and Electromagnetic Research, Zhejiang Provincial Key Laboratory for Sensing Technologies, State Key Laboratory Modern Optical Instrumentation, JORCEP [Joint Research Center of Photonics of the Royal Institute of Technology, Lund University and Zhejiang University], Zijingang campus, Zhejiang University (ZJU), 310058 Hangzhou, *P. R. China*

*** Email for the corresponding author: sailing@jorcep.org**

**Abstract**：

A fluorescence system is developed by using several light emitting diodes (LEDs) with different wavelengths as excitation light sources. The fluorescence detection head consists of multi LED light sources and a multimode fiber for fluorescence collection, where the LEDs and the corresponding filters can be easily chosen to get appropriate excitation wavelengths for different applications. By analyzing fluorescence spectra with the principal component analysis method, the system is utilized in the classification of four types of green tea beverages and two types of black tea beverages. Qualities of the Xihu Longjing tea leaves of different grades, as well as the corresponding liquid tea samples, are studied to further investigate the ability and application of the system in the evaluation of classification/quality of tea and other foods.

## 1. Introduction：

Light-emitting-diodes (LEDs) have been widely studied and used as light sources due to their high energy efficiency in producing monochromatic light as well as the flexibility of changing the spectral region (Bergh, 2004; Pimputkar, et al., 2009). Many analytical systems and commercial instruments have been developed based on the compact and low-cost LEDs (Buah-Bassuah et al., 2008; Dasgupta et al., 2003; Hauser et al., 1995),e.g., LED-based microscopy (Herman et al., 2001). Multi LED light sources or LED arrays have also been used for many applications (Albert et al., 2010; Davitt et al., 2005), since multiple LEDs can deliver plenty of output power while keeping the cost low. On the other hand, multi LED light sources can cover quite a broad spectral range, which can increase the sensitive and applicability of the system (Moe et al., 2005). In fact, the expensive and complicated light sources used in traditional laser-induced fluorescence (LIF) spectroscopy (Anglos et al., 1996; Chappelle et al., 1984) have now been mostly replaced by various LEDs or LED arrays. The LIF method is then often referred to as LED-induced fluorescence spectroscopy, and has found a variety of applications, e.g. teeth inspection (Qin et al., 2007), microscopy (Marais et al., 2008), environmental monitoring (Pan et al., 2003) and ingredient detection (Su et al., 2004). Using of LEDs in these applications makes the systems compact, portable and easily operated. In the present work, an LED-induced fluorescence system with different excitation wavelengths has been developed to evaluate the classification and quality of food.

Tea, a healthy drink originating from China about two thousand years ago, is one of the most popular and largely consumed beverages throughout the world (Cabrera et al., 2006; Weisburger, 1999). Especially, fast tea beverages have become more and more popular. However, it is rather difficult for consumers to evaluate the quality and classification of tea beverages or dried tea leaves on the market due to their similar profile or color. Thus, many different methods have been proposed for tea quality analysis. One method is the high performance liquid chromatography (HPLC) method, which analyzes the chemical components of tea, with the disadvantage of complexity and being time-consuming (Pelillo, 2004; Ying et al., 2005). Near infrared spectroscopy technology (NIR) is also widely used for food quality evaluation (Cen & He, 2007; Huang et al., 2008; Luypaert et al., 2007), and has been proved to be a applicable technology for tea quality and classification assessment ( He et al., 2007; Fu et al., 2013; Luypaert et al., 2003; Ning et al., 2012; Ren et al., 2013; Xu et al., 2013), however, background light can easily interfere the near-infrared light signal and the device used for measurement is expensive. Recently, fluorescence spectroscopy has been found to be very efficient for tea classification and quality assessment (Mei et al., 2012a; Seetohul et al., 2013). However, the laser sources used in the previous work are complex and costly. Therefore, developing a compact and cost-effective fluorescence system by using a multi LED light sources is very attractive and useful for food/tea classification and quality evaluation, as has been discussed above. In the present work, a fluorescence system using multi LEDs covering the spectral range from deep UV to visible light has been developed, and the system has been employed for classification evaluation of tea beverages, and quality assessment of dried tea leaves and the corresponding liquid tea samples.

## 2. Materials and method

### 2.1 Instrumentation

The LED-induced fluorescence system is given in Fig. 1. The compact detection head is a cylindrical cavity made of polished aluminum, including three 5-mm-diameter slots for different LEDs and one vertical elongated hole for a multimode fiber. The LED multiplexing device also has a horizontal flat slit for different filters, which can then be switched by a rotary filter wheel (FW1A, Thorlabs Inc, Newton, USA) according to the excitation wavelengths. The LEDs light source used in the present system can cover a broad spectral range from deep UV to visible light (260 nm to 625 nm), and can be easily installed into the detection head depending upon the applications. However, in the present measurements, only three LEDs of 375 nm, 400 nm and 430 nm peak wavelengths are assembled into the system to induce the fluorescence of the tea beverages and dried tea leaves. The corresponding filters used for eliminating high intensity excitation light in the present work are high-pass filters (OD 4 High Performance Longpass Filter, Edmund Optics Inc, Barrington, USA) with cut-off wavelength at 425 nm, 450 nm, and 525 nm, respectively.

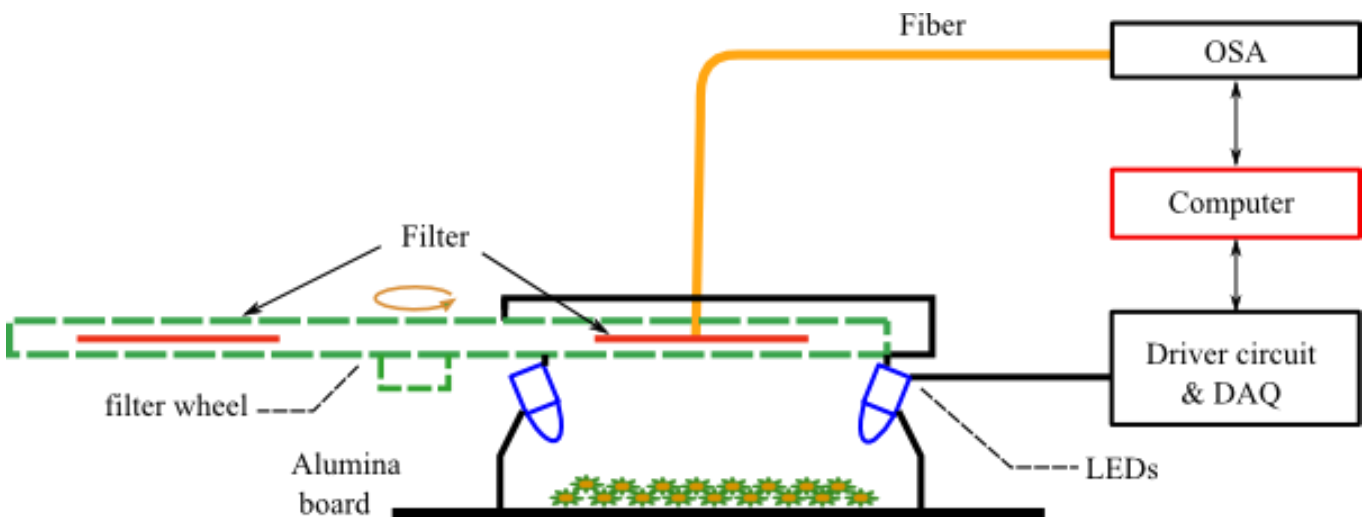

**Fig. 1.** Schematic diagram of the LED-induced fluorescence system

The samples were placed on a black aluminum plate with negligible fluorescence under the swallow hole. The induced fluorescence of the tea samples was first filtered by the high-pass filter to eliminate the direct reflective light at the excitation wavelength, and was then collected by a multimode fiber with a core diameter of 1.0 mm. The induced fluorescence was then measured by an optical spectrum analyzer (OSA, USB 2000, Ocean Optics Inc, Dunedin, USA) for analysis. The emission intensities of the LEDs were controlled by a data acquisition card (DAQ, National Instruments, USB 6008) through a driving circuit. The LEDs could be illuminated one-by-one, and synchronized with the OSA through a LABVIEW (Labview 2010, National Instrument Co Inc, Austin, USA) based program during the measurement. Of course, a specific LED can be turned on if the fluorescence spectrum at this particular wavelength is desired. Finally, the fluorescence spectra were recorded in the computer

## 2.2 Tea samples

Four types of green tea beverages – Wow Haha tea beverage (WG), Uni-President tea beverage (UG), Nestle tea beverage (NG) and Tingyi tea beverage (TG) - bought from the supermarket, were studied in our work. They are made by nature tea leaves with some additives. It is difficult to distinguish these tea beverages due to their similar color and taste. Two kinds of black tea beverages with similar colors, Nestle tea beverage (NB) and Oriental Leaf black tea beverage (OLB), were also examined to verify the classification ability of this system. The pictures of the tea beverages are shown in Fig. 2.

Xihu Longjing tea samples comprising seven different grades, provided and evaluated by a Longjing tea company in Hangzhou, Zhejiang Province, China, were also investigated by this method. The seven groups of tea leaves were collected from Longjingcun, a local tea plantation and stored in air-tight packaging. Their qualities are marked from grade one to grade seven, with higher grade number denoting better quality and higher price. Of course, the grades evaluated by the experts might not exactly indicate the real quality of the tea samples. The corresponding liquid tea samples were prepared by mixing 100 mL hot water at 100 $^{O}$C with 1.0 g dried Xihu Longjing tea leaves for 30 min. The tea leaves were filtered out before the measurements.

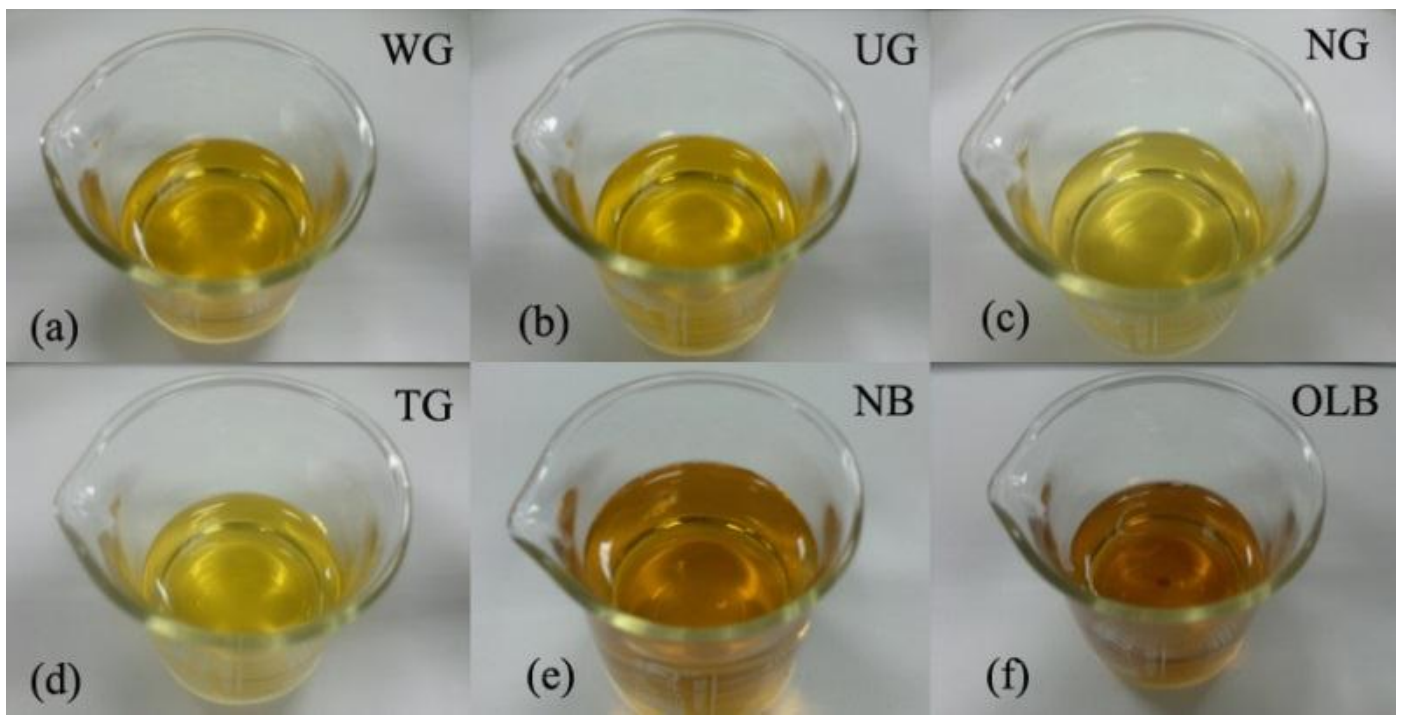

**Fig. 2.** Pictures of six types of tea beverages: (a-f) correspond to Wow Haha green tea (WG), Uni-President green tea (UG), Nestle green tea (NG), Tingyi green tea (TG), Nestle black tea (NB) and Oriental Leaf black tea (OLB).

## 2.3 Analysis method

The collected spectra are analyzed by principal components analysis (PCA) based on singular value decomposition (SVD), which can significantly reduce the dimensions of spectra data by using the so-called principal components (PCs); the original fluorescence spectra could then be represented by several main PCs. Each spectrum is removed in turn, and the remaining spectra data are used to build the predictive model based on linear discriminant analysis (LDA) (Mei et al., 2012a). If ***L*** is the predefined matrix of the quality/classification of tea given by tea experts, matrix ***P*** is the contribution of *m* PCs to their spectra except the one left out (denoted as the $j^{th}$ sample). The linear coefficient matrix ***S*** can be determined by:

$$S_{0:m;1:k} = L_{i=1:n,i\neq j;1:k} / P_{i=1:n,i\neq j;0:m} \tag{1}$$

Here, *n* is the total number of data samples, *k* is the number of variables for each sample, and *i* expresses the $i^{th}$ sample of the *n* samples. The evaluated grade or classification value for the corresponding fluorescence spectrum could be obtained by：

$$L_{j;1:k} = P_{j;0:m} \bullet S_{0:m;1:k} \tag{2}$$

The same evaluation procedure was performed for all the fluorescence spectra, and the classification/quality of the tea samples was obtained through a program based on MATLAB (Matlab R2008b, The MathWorks, Natick, USA). However, sometimes several-time measurements would be performed for one sample, thus the evaluated classification/quality for each sample was retrieved by averaging the values calculated from all the recorded fluorescence spectra. In this experiment, each beverage was measured twenty times for each LED source to enhance the accuracy of classification evaluation. Fifteen data of each group were selected as prediction set to build up a model and the other five data were used as validation set to further test the model. Five spectra were recorded for each sample to obtain the average evaluation result when measuring tea leaves and liquid tea.

According to Eq. (1), the classification matrix ***L*** is defined as:

$$L_{n,t} = \begin{cases} 1 & n \subset class\, t \\ 0 & n \not\subset class\, t \end{cases} \quad n = 1, 2 \ldots N \tag{3}$$

where the classification of tea beverages is indicated by *t*=1, 2…*T*, *T* represents the categories of beverages and *N* denotes the number of the fluorescence spectra, the size of matrix ***L*** is 80×4 since *N*=80.

A classification index *Q*, defined in Eq. (4), describes the discrimination index for each classification.

$$Q = | u_{L_n \subset t} - u_{L_n \not\subset t} | / (\sigma_{L_n \subset t} + \sigma_{L_n \not\subset t}) \tag{4}$$

where $u_{L_n \subset t}$ and $\sigma_{L_n \subset t}$ represent the mean and standard deviation of the evaluated classification values of class *t*, respectively, while $u_{L_n \not\subset t}$ and $\sigma_{L_n \not\subset t}$ are the mean value and standard deviation of other classifications. Obviously, class t with more uniform data and larger distinction from other classes in the measurement could obtain a higher $Q$ value.

## 3. Results and discussion

### 3.1 Classification for four different green tea beverages

Four kinds of green tea beverages samples were measured each for 20 times by the LED-induced fluorescence system. The averaged fluorescence spectrum for each excitation wavelength is normalized to the highest fluorescence peak which is typically around 500nm, as shown in Fig. 3. However, for the classification analysis, all 20 recordings of each sample consist of the whole fluorescence spectra database, and thus there are 80 fluorescence spectra. The reflectance before 450 nm cannot be fully eliminated by the corresponding long pass filter, so the analysis is performed with the fluorescence spectra above 450 nm to avoid the interference from the emission of the LEDs.

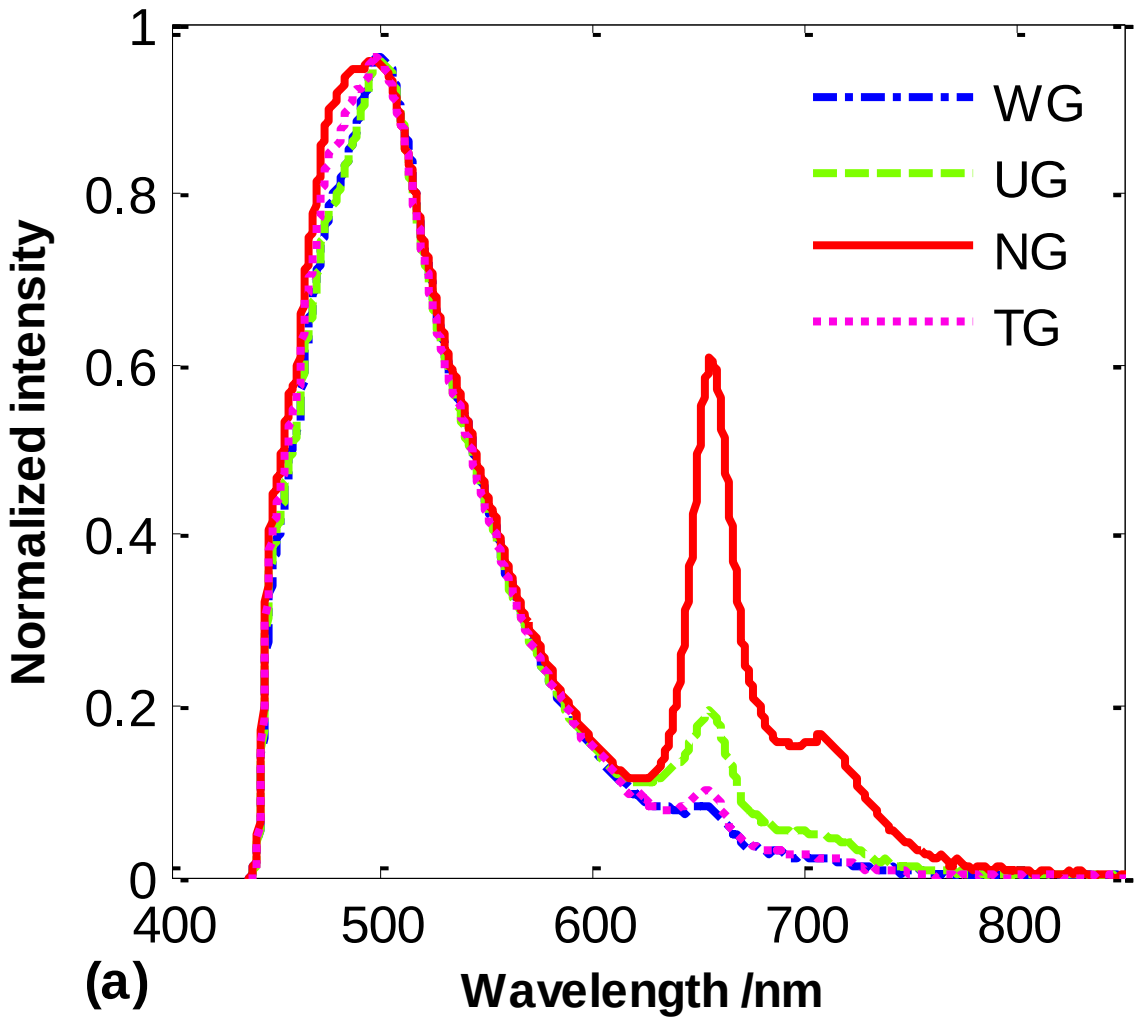

(a)

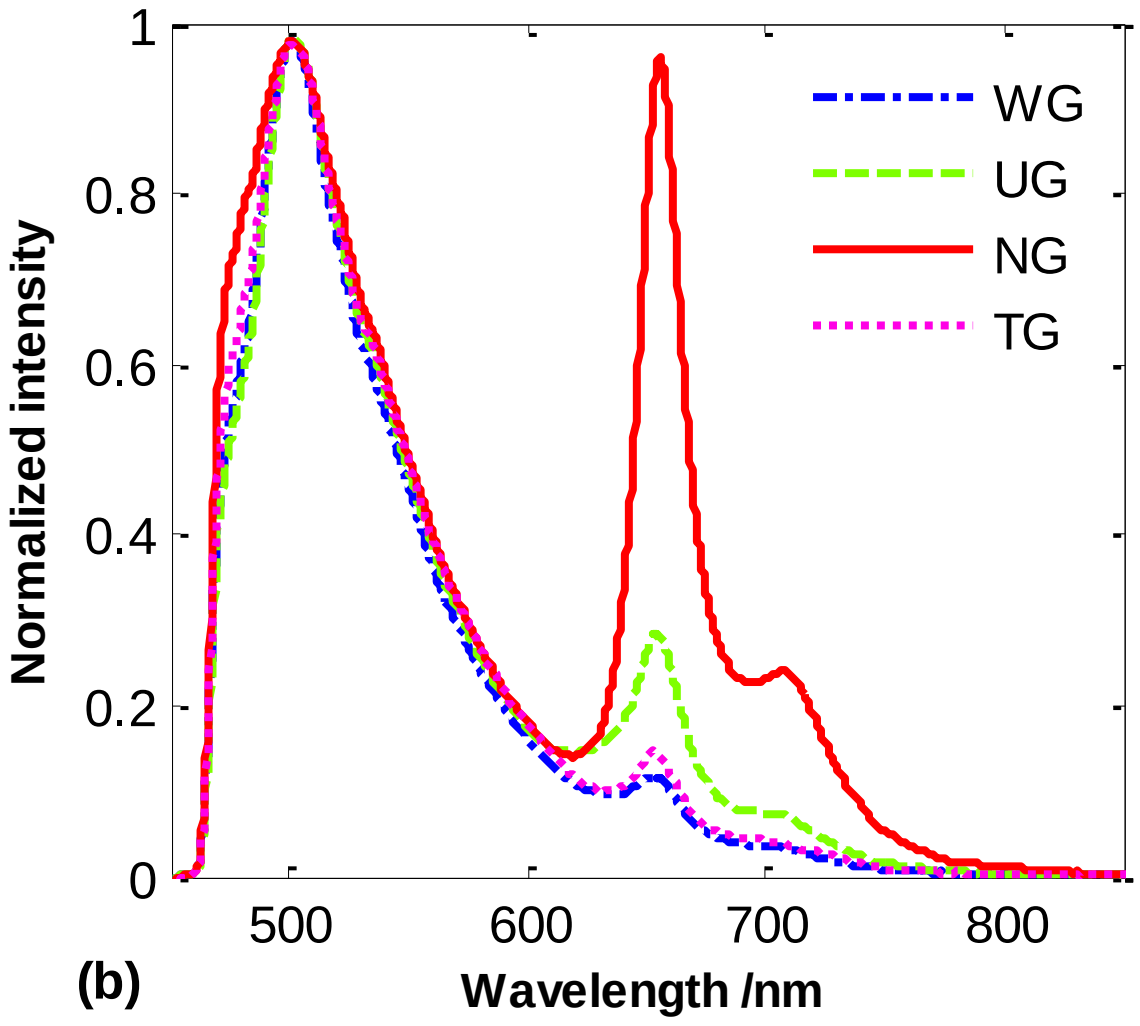

(b)

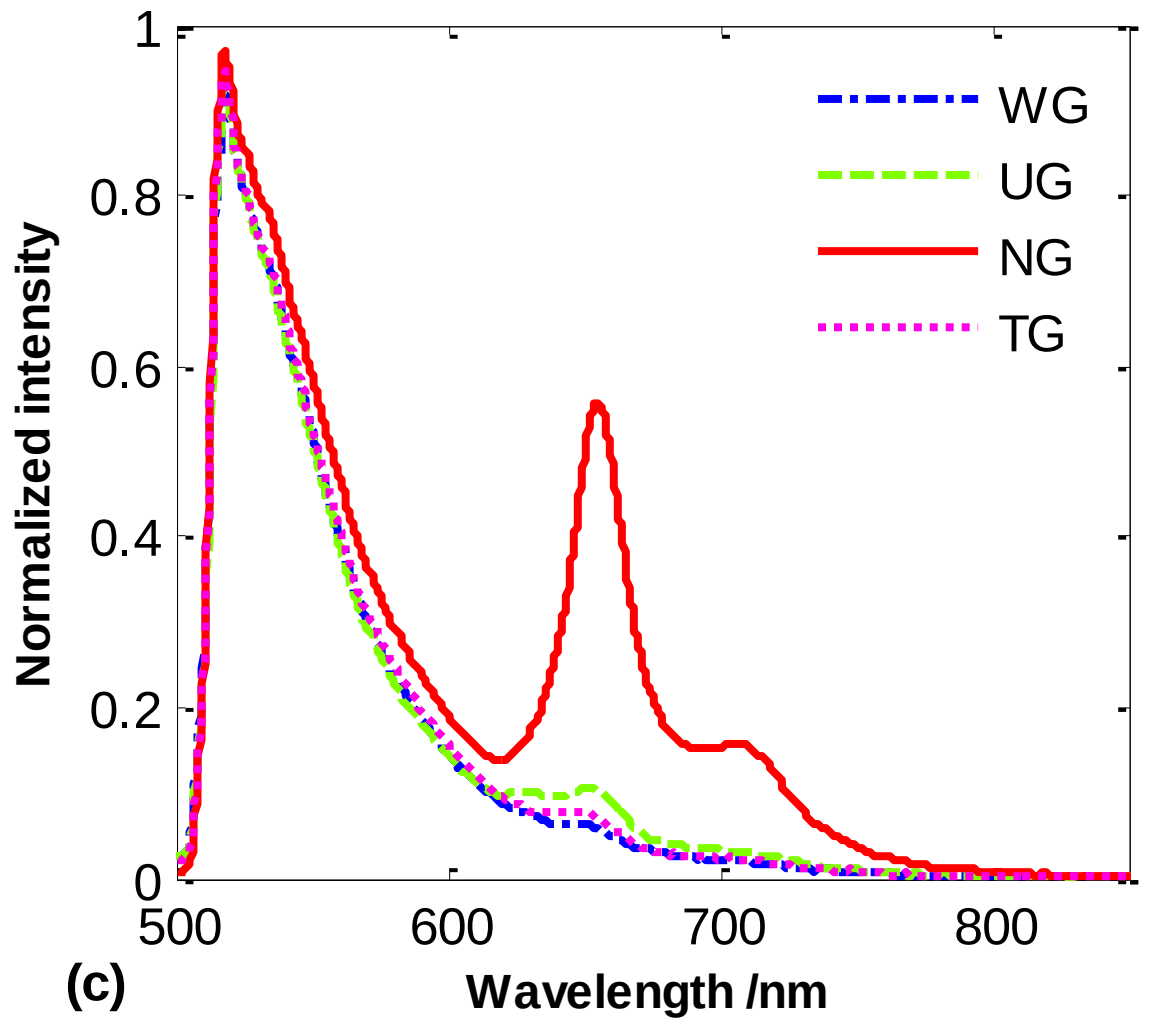

**Fig. 3.** Averaged fluorescence spectra induced by 375 nm (a), 400 nm (b), and 430 nm (c) LED for four different green tea beverages.

As shown in Fig. 3, the differences of the spectra around 660nm are mainly due to different chlorophyll contents in the samples. However, the profile of the spectra from 620nm to 770nm is a synthetic reflection of chlorophyll and polyphenols, catechins and theaflavins. The peak around 520nm is determined by some organics such as protein. The categories of the four green tea beverages might be seen from the fluorescence spectra directly. However, from a methodological point of view, it is still worthwhile to use the statistical method such as PCA analysis to analyze the whole spectra from 500nm to 800nm for the evaluation of the classification, since in this way a clear and convincing classification result could be displayed. In the present work, only $m$=6 principal components are selected to represent the whole fluorescence spectra.

The evaluated values/scores, retrieved from Eq. (1) to Eq. (3), represent how well the corresponding tea samples match with a given classification, where a score of zero means that the sample does not belong to the specific classification. As an example, the evaluated classification values using the 400nm LED as excitation light source are given in Fig. 4. The $Q$ values for four tea beverages with three LED sources are given in Table 1, which is calculated from all the twenty evaluation results of both the prediction data set and validation data set. According to our experience, it is not easy to distinguish one tea beverage from another if its $Q$ value is too low, eg., less than 1.4. Comparing the $Q$ values of three excitation wavelengths in Tab.1, it is obvious that 400nm case is best, because all $Q$ values of which are large enough to be distinguished as shown in Fig.4. As for 375nm and 430nm cases, however, they are relatively difficult to distinguish WG from others. It indicates that the 400 nm LED source should be used for this application. However, LED sources at other wavelengths might still be useful for some other types of tea beverages or other food applications.

**Table 1**. Evaluated classification results for four green tea beverages with different LEDs as the excitation light source. The $Q$ values indicate the discrimination indices corresponding to WG, UG, NG and TG, respectively.

| Excitation wavelength(nm) | Evaluated classification | | | |
| --- | --- | --- | --- | --- |
| | $Q_1$ | $Q_2$ | $Q_3$ | $Q_4$ |
| 375 | 1.36 | 4.76 | 6.51 | 1.96 |
| 400 | 2.48 | 3.19 | 11.8 | 2.17 |
| 430 | 0.69 | 2.90 | 7.18 | 1.76 |

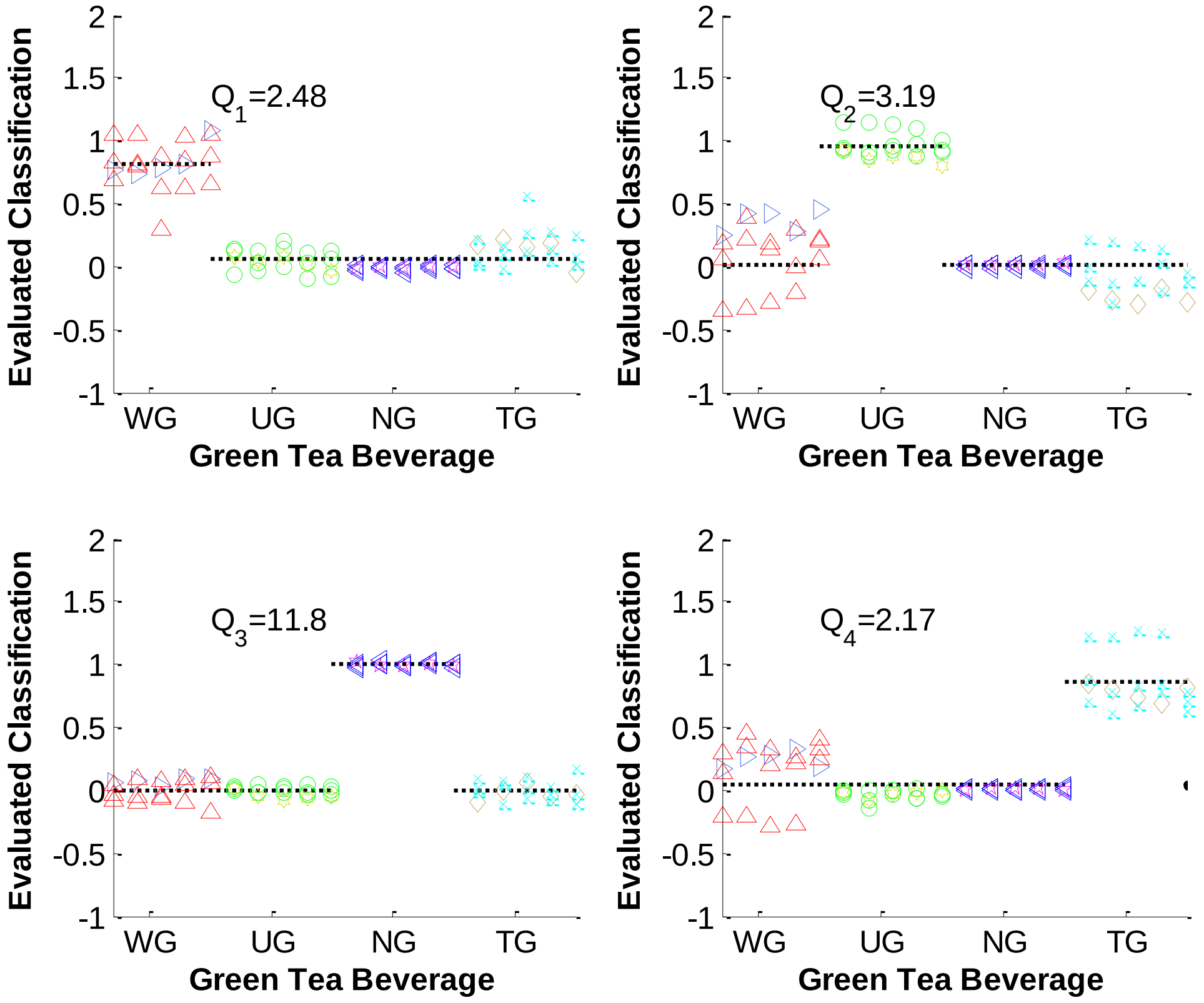

**Fig. 4.** Predicted classification values of prediction data set for four kinds of green tea beverages: WG (△), UG (○), NG(◁), TG (×), and of validation data set for the same tea beverages: WG (▷), UG(✡), NG (☆), and TG (◇) with a 400 nm LED as the excitation source. The dotted lines predict the mean evaluated value, and $Q_{1...4}$ stands for the discrimination degree of each beverage.

## 3.2 Classification for six types of tea beverages

The fluorescence spectra of two types of black tea beverages were also measured and added into the fluorescence spectra database to further verify the classification ability of the LED fluorescence system. Fig. 5 gives the spectra for the green and black tea beverages induced by the 400 nm LED, since the green tea beverages can easily be distinguished at this wavelength. From Fig. 5, broader visible fluorescence spectra for the black tea beverages are observed, which will possibly mitigate the difficulties of distinguishing green and black tea beverages. Using the same evaluation procedure, the $Q$ values for six different types of tea beverages were obtained – 1.44 (WG), 6.49 (UG), 15.6 (NG), 1.64 (TG), 8.47 (NB), and 7.27 (OLB). Comparing this result with the results given in Table 1, the $Q$ values of the green tea beverages WG and TG are smaller, indicating increased difficulties of classification when adding the fluorescence spectra of the black tea beverages into the spectral database. One possible reason is that the matrix $\boldsymbol{P}$ (as described in section 2.3) changed a lot when two new spectra were added for analysis, and could be eliminated as a complication when a much larger spectral database is established. However, generally speaking, the six-type tea beverages are all able to be distinguished.

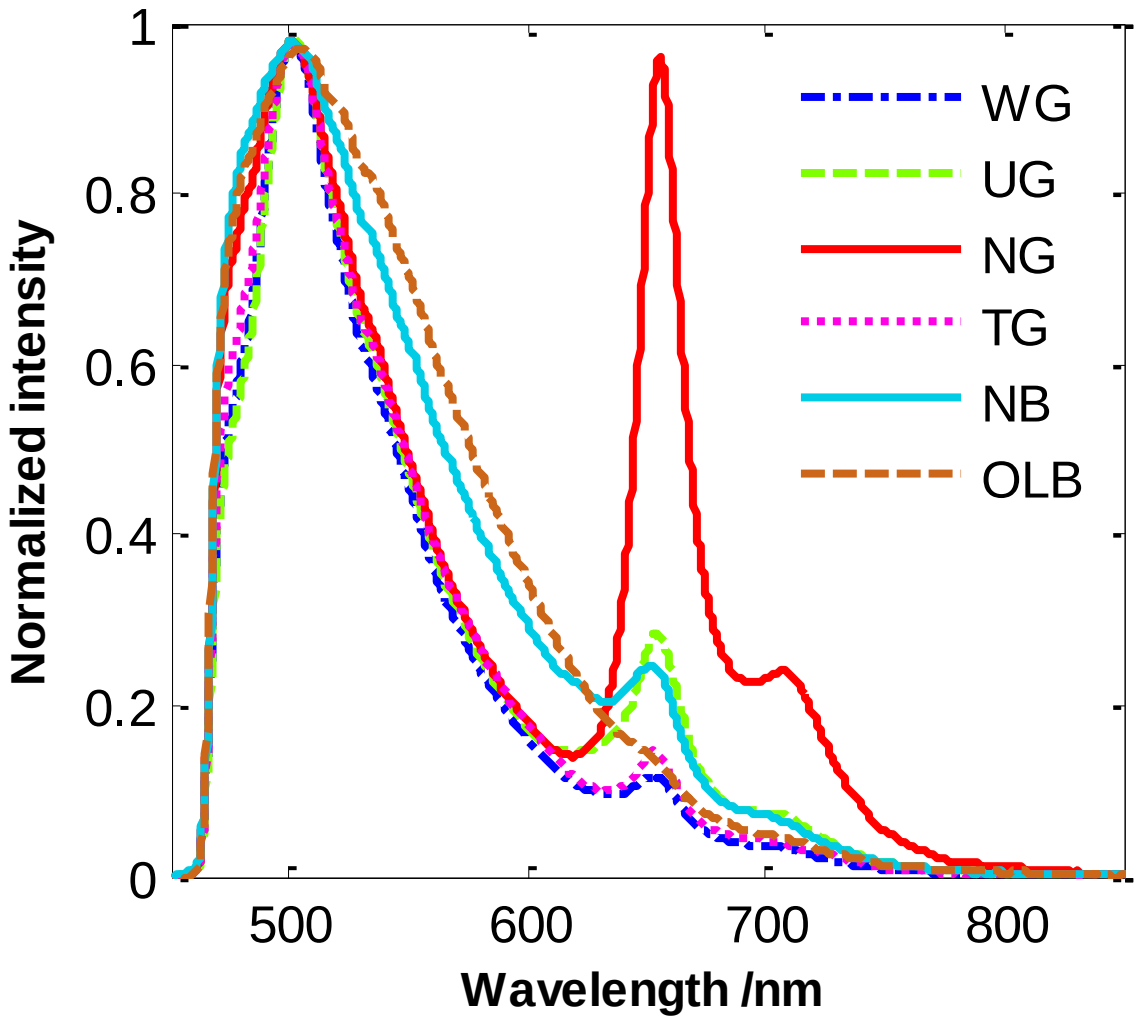

**Fig. 5.** Averaged fluorescence spectra of six tea beverages, including four types of green tea beverages (WG, UG, NG, TG) and two types of black tea beverages (NG and OLB), induced by a 400 nm LED.

### 3.3 Quality assessment for Xihu Longjing tea

Seven Xihu Longjing tea samples, which cannot be distinguished via visual observation, were measured using the LED fluorescence system. In order to avoid non-uniformity when performing the measurements on the dried tea leaves, the corresponding liquid tea samples were measured the same way as the tea beverages. After the PCA analysis, the fluorescence spectra induced by the 400 nm LED were found to be very efficient for tea quality/grade evaluation. The averaged fluorescence spectra for the dried tea leaves and liquid tea from grade one to grade seven are shown in Fig. 6.

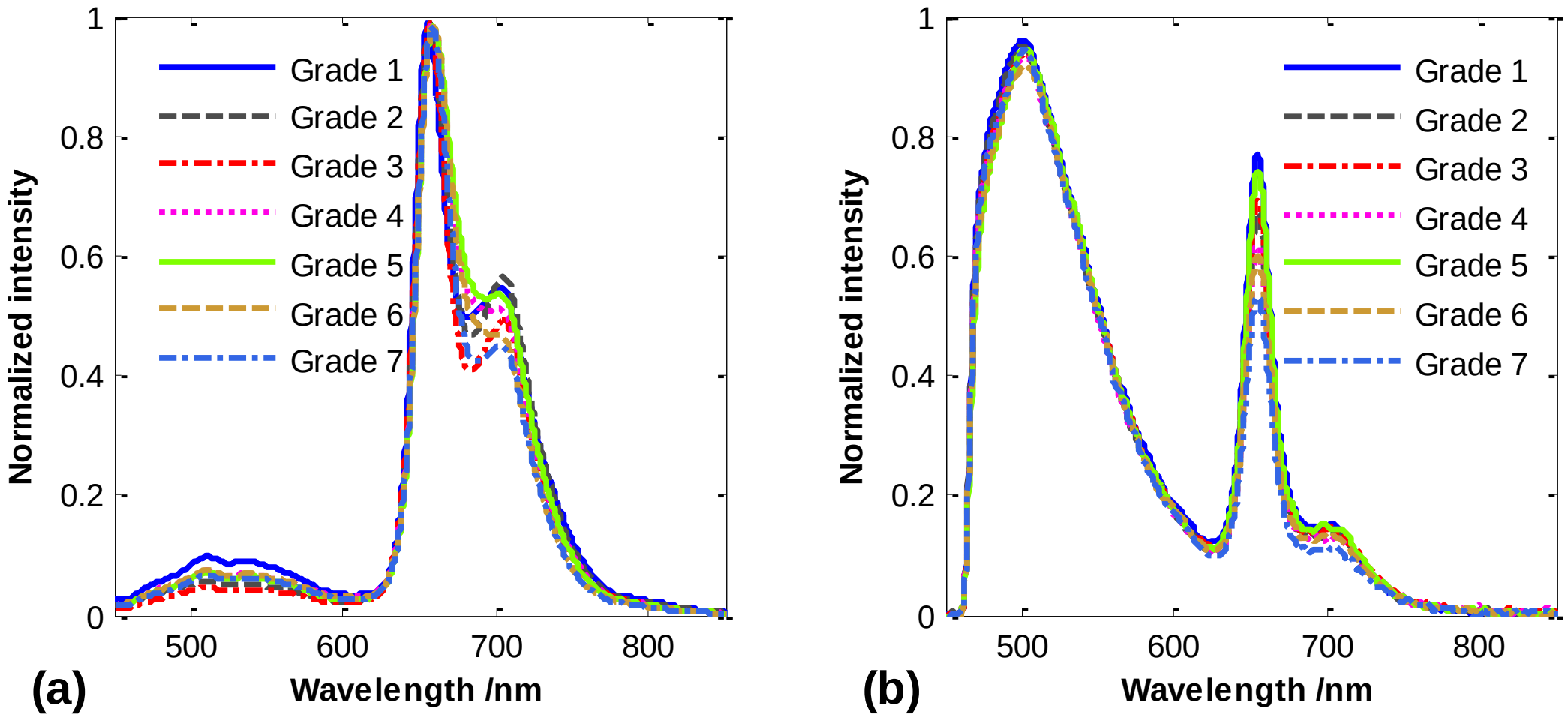

**Fig. 6.** Averaged fluorescence spectra of Xihu Longjing tea leaves (a) and liquid tea (b) with different grades from one to seven when measured by a 400 nm LED.

The spectral differences of the seven samples are not very obvious, which means that the qualities/grades could not be directly determined from the spectra. The PCA analysis is applied with a predefined matrix ***L*** given as

$$L = [1\ 2\ 3\ 4\ 5\ 6\ 7] \qquad (5)$$

The evaluated tea grades are obtained according to Eqs. (1), (2) and (5), the results both for the dried tea leaves and liquid tea samples are given in Fig. 7. Comparing with the negative results for the dried Longjing tea leaves using 355 nm excitation wavelength presented in Mei et al. (2012b), the grades assessed through the LED fluorescence system agree well with grades of tea samples decided by experts, with a correlation factor of 0.99 and 0.98 for dried tea leaves and liquid tea samples, respectively. The multi LEDs obviously provide more possibility for quality evaluation. The good evaluated results with high correlation values and similar trends for the dried tea leaves and corresponding liquid tea samples indicate the present method as a powerful tool for tea quality assessment.

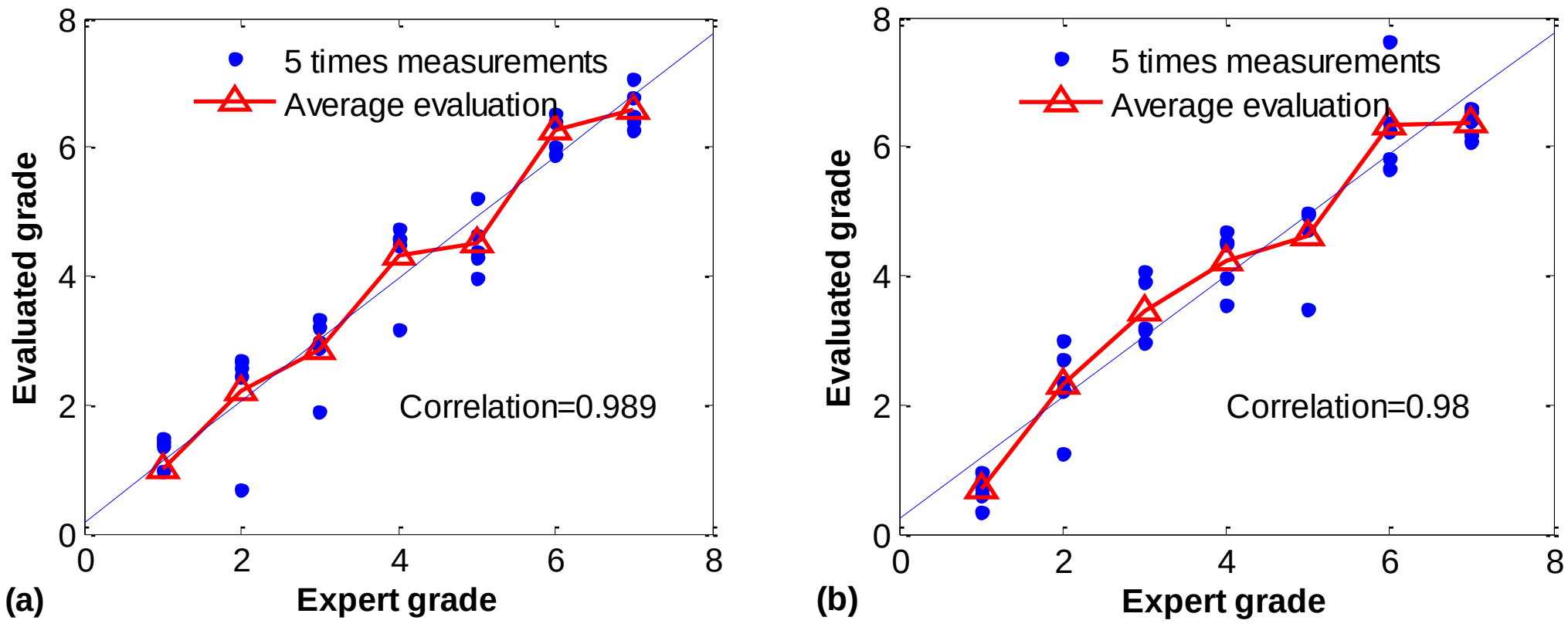

**Fig. 7.** Grade assessment of Xihu Longjing tea leaves (a) and liquid tea (b) with seven different grades. The blue dots are the evaluated grade calculated from each measured spectrum and the red triangles indicate the averaged grade for seven groups. The red line and the blue line represent the correlation and linear fitting of evaluated grades to expert grades, respectively.

The promising results indicate that the LED fluorescence system has great potential for quality assessment of both the dried tea leaves and liquid tea samples, although their fluorescence spectra are quite different. In summary, depending upon the applications, either dried tea leaves or the corresponding liquid tea samples can be measured, and similar results are expected.

## 4. Conclusion

In this study, an LED-induced fluorescence system has been developed with a multi LEDs light source, which is low-cost, compact and convenient for use. Various application prospects related to food quality/classification can be further exploited. According to the present experimental results, the ability of the system for tea beverages classification has been demonstrated. On the other hand, although the completely exact grade evaluation has not been achieved, the rough estimation of tea quality to satisfy the requirement of consumers is verified using this system.

In the present measurements, only three LEDs are installed and the LED with a 400 nm peak wavelength has been found to be very efficient for the classification and quality assessment of the beverages and dried tea leaves. More LEDs covering the wavelengths from the deep UV to visible region could be mounted on the detection head, so that the sample could be measured with LEDs at different wavelengths. Generally, the LED-induced fluorescence system could be further improved and developed into a portable device used by consumers for classification and quality assessment, or applied in the industries for online monitoring.

**Acknowledgements**

The authors greatly acknowledge the help from Dennis Alp and Samuel Modée of KTH during the measurement. This work was partially supported by the Science and Technology Department of Zhejiang Province (2010R50007).